
----------
X-Sun-Data-Type: default
X-Sun-Data-Description: default
X-Sun-Data-Name: preprint1.tex
X-Sun-Content-Lines: 863

%

\def\ga{ \gamma }
\def\mpe{ m_p/m_e }
\def\flu{\ergs\cmsqi}
\def\Suxg{ S_{u,x,\gamma}}
\def\ka{ \kappa}
\def\la{ \lambda}

\def\Kpc{{\rm\,kpc}}

\def\cm{{\rm\,cm}}

\def\ga{ \gamma}
\def\etal{{\it et~al.}}

\def\halfspace{\baselineskip=12pt plus .1pt}

\def\papersize{\magnification=1200}  

\font\tfont=cmmib10
\newfam\vecfam

\textfont\vecfam=\tfont \scriptfont\vecfam=\seveni
\scriptscriptfont\vecfam=\fivei

\parindent=40pt
\settabs 7 \columns
\tolerance=1600
\parskip 1ex
\def\foolit{\ifnum\pageno > 1 \number\pageno\fi}
\raggedbottom
\def\frac#1#2{{#1\over#2}}
\def\Mesz{M\'esz\'aros\ }
\def\Pacz{Paczy\'nski\ }

\def\ctl{\centerline}

\def\ref{\par \noindent \hangindent=2pc \hangafter=1 }
\def\etal{{\it et~al.\ }}
\def\mathnew{\mathsurround=0pt}
\def\simov#1#2{\lower .5pt\vbox{\baselineskip0pt \lineskip-.5pt
	\ialign{$\mathnew#1\hfil##\hfil$\crcr#2\crcr\sim\crcr}}}
\def\simg{\mathrel{\mathpalette\simov >}}
\def\siml{\mathrel{\mathpalette\simov <}}
\def\lambdabar{\mathrel{\lower 1pt\hbox{$\mathchar'26$}\mkern-9mu
        \hbox{$\lambda$}}}

\def\ssk{\vskip 1ex\noindent}

\def\bsk{\vskip 3ex\noindent}

%
\def\cm{~\rm{cm}}

\def\s{~\rm{s}}

\def\cmsqi{~ {\rm cm}^{-2} }
\def\cmcui{~ {\rm cm}^{-3} }

\def\eV{~\rm{eV}}

\def\erg{~\rm{ergs}}
\def\ergs{~\rm{ergs}}

\papersize
\halfspace
%
%
\ctl{\bf GAMMA-RAY BURSTS: MULTIWAVEBAND SPECTRAL PREDICTIONS}
\ctl{\bf FOR BLAST WAVE MODELS}
\bsk
\ctl{ P. \Mesz$^{1}$ and M.J. Rees$^{2}$}
\bsk
\ctl{$^1$~ Pennsylvania State University, 525 Davey Lab, University Park, PA
16803}
\ctl{$^2$~ Institute of Astronomy, Madingley Road, Cambridge CB3 OHA, England}
\bsk
\bsk
\ctl{ Submitted to Ap.J.(Letters): ~8/13/93~~~;~ accepted:~9/8/93 }
\bsk
%
%
\ctl{\bf Abstract }
\bsk
In almost any scenario for 'cosmological' gamma-ray bursts (and in many models
where they originate in our own Galaxy), the initial energy density is so large
that the resulting relativistic plasma expands with $v\sim c$ producing a blast
wave ahead of it and a reverse shock moving into the ejecta, as it ploughs into
the external medium.  We evaluate the radiation expected from these shocks,
for
both cosmological and galactic bursts, for various assumptions about the
strength of the magnetic field and the particle acceleration mechanisms in the
shocks. The spectra are evaluated over the whole range from the IR to $>$ GeV,
and are compared with the variety of spectral behavior reported by BATSE,
and with the X-ray and optical constraints.  For bursts of duration $\simg 1\s$
acceptable $\gamma$-ray spectra and $L_x/L_\gamma$ ratios are readily obtained
for 'cosmological' models. Blast waves in galactic models can produce
bursts of similar gamma-ray fluence and duration, but they violate the X-ray
paucity constraint, except for the shorter bursts ($\siml 1\s$).  We discuss
the prospects for using O/UV and X-ray observations to discriminate among
alternative models.
\bsk
%
\ctl{\bf 1.~ Introduction}
\bsk
Gamma-ray bursters (GRBs) emit most of their energy above $\sim 0.5$ MeV (e.g.
Band, \etal, 1993) and have not so far been convincingly detected at energies
below a few KeV, despite intensive searches (e.g. Hartmann, 1993). Relatively
large error boxes and the lack of good calibrations have provided only
tentative
upper limits from archival searches (e.g. Hudec, \etal, 1992), while active
optical monitoring has been carried out only sporadically and on a modest scale
(e.g. Moskalenko, \etal, 1992). Quasi-simultaneous X-ray emission has been
detected in only a few objects (e.g. Murakami, \etal, 1992, Boer, \etal, 1992),
and limits have been set on the X-ray to $\ga$-ray luminosity ratio (X-ray
paucity constraint) of $\siml \hbox{few}~10^{-2}$, e.g. Laros, \etal, 1984;
Hurley, 1989).
However, a new era of multiwaveband observing capability is expected to be
ushered in with
the launch in a few years of HETE (Ricker, 1992), which will carry broad-band
gamma-ray,
X-ray and UV detectors. This should improve the chances of finding
counterparts and provide more accurate low frequency fluxes or upper limits.

If GRBs are at cosmological distances, the luminosity is so high that the
dominant radiation mechanism would involve relativistically outflowing plasma.
Such an outflow is expected even in galactic models (e.g. those involving
violent disturbances of neutron star magnetospheres). In the latter case,
Begelman, \etal (1993) propose that only the bursts classed as
``short/variable"
(Kouveliotou, \etal, 1993; Lamb, Graziani \& Smith, 1993) come from the
magnetosphere,
and attribute the commoner "long/smooth" bursts to interaction of a
relativistic
outflow with external matter.

For relativistic blasts, the apparent diameter of the emitting region is larger
by
at least a factor $\Gamma$ (the bulk Lorentz factor) than in the corresponding
nonrelativistic case. Consequently there is much less likelihood of strong
suppression of
the radiation down to optical frequencies due to self-absorption.  For this
reason, and
also because the radiation mechanisms are more straightforward than in some
alternative models (no ultra strong magnetic fields, etc.), any O-UV/ X-ray
measurements or upper limits would have relatively straightforward implications
for this promising class of models.
\bsk
\ctl{\bf 2.~ Shock Structure and Magnetic Fields}
\bsk
The relativistic ejecta may be either a radiation-pair fireball with some
baryon
loading, or  could  be predominantly Poynting flux if the initial event were
strongly magnetically dominated. In either case, sweeping-up of the external
medium will create a highly relativistic  shock that can accelerate particles
to relativistic energies; the reverse shock propagating back into the ejecta
will also generally be strong and relativistic. Even if the magnetic fields
already present in the ejecta or in the external medium are weak, they may be
amplified up to equipartition values in the shocks (as seems to happen in
supernova remnants and radio sources,  probably because shock instabilities,
e.g. Ryu and Vishniac, 1992, lead to turbulent field growth).  The observable
burst timescale $t_b$ is determined by the external density $n_o$ and the total
burst energy $E_o$, if the initial energy deposition is impulsive (occurs on
$\Delta t \ll t_b$); if $\Delta t$ is longer, it is determined by the energy
release
mechanism itself (Rees and \Mesz 1992).

The radiative efficiency of a GRB shock is a function of the bulk Lorentz
factor,
the magnetic field strength, the particle acceleration mechanism and the
density
of the external medium. Various combinations of these elements can lead to
shocks
with different spectra and radiation efficiencies (\Mesz and Rees, 1993, Katz,
1993).
The minimum electron (random) Lorentz factor behind the blast wave is expected
to be $\gamma_m =\kappa \Gamma$, where $\Gamma$ is the bulk Lorentz factor of
the
shock and $1\siml \kappa\siml \mpe$ (the latter ocurring if energy is shared
between electrons and protons in the shock). Diffusive shock acceleration is
expected to produce a power law electron spectrum above $\gamma_m$, whose slope
is typically -2 to -3 (e.g. Blandford and Eichler, 1987).

The properties of the reverse shock depend on the nature of the fireball. In
the
case of a baryon-loaded fireball the sound speed in the ejecta may be low, and
the reverse shock will be strong and at least mildly relativistic. If the
fireball
is magnetically dominated (almost completely unloaded, e.g. Narayan, \Pacz and
Piran, 1992), it will have a large outward Lorentz factor even relative to the
frame of the contact discontinuity. The reverse shock is then strong, the field
in the shocked region providing most of the pressure that drives the blast
wave.
Particles (either entrained, or neutrals  swept up from the external medium),
will
be accelerated, rather as in the Crab's wisps (e.g. Hoshino, \etal, 1992).
\bsk
\ctl{\bf 3.~ Spectral Components}
\bsk
The radiative efficiency, and the spectrum, depend on the physical
nature of the outflow material, the efficiency of field growth and particle
acceleration behind the shocks, and the degree of mixing across the contact
discontinuity.  The energy bands at which the photons appear depend on the
electron Lorentz factors $\ga$ produced in the respective shocks.  If an
adequate magnetic field is present, either through shock amplification or
(in the case of the reverse shock) because it is frozen into the ejecta, each
shock contributes a synchrotron component, and a higher energy component due to
IC scattering of the synchrotron photons by the same electrons (\Mesz, Laguna
and Rees, 1993).

If both reverse shock and blast wave acceleration are efficient, a third
additional IC component arises, due to upscattering of reverse synchrotron
photons by blast wave electrons. (We neglect IC scattering in the reverse
shock of any photons originating in the blast wave, which is a relatively
weaker
photon source, and also second-order IC, which is likely to occur in the
Klein-Nishina regime).

If there is no frozen-in field, and only the blast wave leads to field growth,
a two component spectrum arises. On the other hand, if  the only strong field
is associated with the reverse shock, a three component spectrum is expected:
two from the reverse-shocked ejecta, plus a component due to IC scattering of
this radiation by the high-energy electrons that are, in all cases, accelerated
behind the ultrarelativistc blast wave . Finally, if substantial field growth
occurs in both shocks (or the reverse shock is dominated by frozen-in fields
and
growth occurs in the blast wave), and acceleration is efficient in both, there
will be five components altogether.

The energy slope of the synchrotron spectrum is $-(p-1)/2$ above frequencies
$\nu_m\sim 10^6 B\gamma_m^2$, where $-p$ is the electron power-law index above
$\gamma_m$. We took $p$ to be around -3, giving a photon energy index about
$-1$
above $\nu_m(\ga_m)$. For a single value of $\ga_m$, the photon energy spectrum
below $\nu_m(\ga_m)$ would be expected to have a slope $+1/3$. (Synchrotron
self-absorption is included, but usually occurs below the UV range for the
parameters
of these models). The corresponding
power per decade slopes below and above the break $\nu_m$ would be 4/3 and 0,
close to the observed ``fiducial" values 1 and 0 deduced for those objects
where
a break is required (Schaefer, \etal, 1993; but cf Band \etal 1993)). However,
$\ga_m$ may not be uniform throughout different regions of the shock, which
could
easily lead to a flattening of the slope below the break. We therefore  assume,
in accord with the data, a fiducial power per decade with slope 1 (energy slope
0)
below the break. Thus the power per decade composite spectrum of GRBs, whether
galactic or extragalactic, is made up of between two to five subcomponents with
slopes roughly 1 (0) below (above) break energies determined by the comoving
mean values of $B,~\ga_m$ (up to two synchrotron, up to three IC breaks
and components).

The importance of each spectral component is found by estimating the
corresponding radiative efficiency, based on the values of $\ga_m,~p$ and
$u_B,~u_r$ (the comoving magnetic and radiation density) in each shock. The
efficiency of each radiation mechanism in a particular shock is $j$ is
$e_j=t_j^{-1}/(t_j^{-1}+t_i^{-1}+t_{ex}^{-1})$, where $t_i$ is the harmonic
sum of the other competing radiative mechanisms and $t_{ex}$ is the expansion
time in the comoving frame. The total available energy in each shock is a
fraction of the kinetic energy of the ejected material ($\sim E_o$, the initial
energy liberated). This is shared approximately equally between both shocks,
since they are in pressure equilibrium, e.g. Rees and \Mesz, 1992. The total
fluence is then determined by multiplying this fraction by the efficiency and
by the total bolometric fluence potentially available, $S_b=E_o/(4\pi\theta^2
D^2) \erg\cmsqi$, where $\theta$ is the possible beaming angle and $D$ is the
distance.
\bsk
\ctl{\bf 4.~ O/UV, X, and $\gamma$ Fluences}
\bsk
The predicted spectrum may in principle show up to five breaks. In most
cases, however, the net spectra are dominated by not more than three
distinguishable components, the break energies depending on the density,
magnetic field, etc. If the main break is outside all three observational
bands or windows considered, the  fluence in the $u,~x$ windows relates to
that  in the $\ga$ window as $S_u/S_\ga=(E_u/E_\ga)^{\alpha},~S_x/S_\ga =
(E_x/E_\ga)^{\alpha}$, where $(E_u,E_x,E_\ga)\sim (5\eV,10^3\eV,5\times 10^5
\eV)$ are the appropriate energies of the bands. The fiducial value of $\alpha$
is 1 for a high-energy break and 0 for a low-energy break. In general,
especially
if two or three breaks dominate the total spectrum, these spectral breaks come
between or within the bands, and the ratios are more complicated.

We have performed calculations for a range of $E_o,~D$ and $\theta$
corresponding
to both extragalactic and galactic models, for various assumptions about the
magnetic field generation and frozen-in component, as well as about the
electron
acceleration parameters and the strength of the shock $\eta=E_o/M_oc^2$ for
matter-dominated shocks or $\Gamma$ for magnetic dominated shocks.
In the cosmological cases we took $D=10^{28}\cm$ and in the galactic cases
$D=1\Kpc$.
A modest variation of $\eta$ by one order of magnitude results in a range of
burst durations $0.1\s \siml t_b\siml 10^3\s$, while the burst diversity is
easily explained by variability in the density of the external medium
encountered
and/or cooling timescale substructure induced by magnetic inhomogeneities in
the
shocks caused by instabilities or turbulence. We report here some results for
typical parameter values; further details will be discussed elsewhere.

If turbulent field growth leads to equipartition fields in both the reverse
shock
and the blast wave , the models show at most three dominant spectral
components.
For cosmological models with $E_o=10^{51}\ergs,~D=10^{28}\cm$ and $\theta=
10^{-1}$, with $\eta=10^3$ and external density $n_o=1\cmcui$, the burst lasts
5 s, and for equipartition fields one gets fluences of $\log\Suxg\sim
(-5.9,-4.9,-4.7),~(-9.3,-6.8,-6.7)$ in $\flu$, for $\kappa=1,~10^3$. Similar
models at galactic disk distances with $E_o=10^{39}\erg,~D=1\Kpc$ and
$\theta=10^{-1}$, $\eta=10^2$ and $n_o=1 \cmcui$ give bursts that last
5 s with fluences $\log\Suxg\sim (-6.5,-6.5,-6.5),~(-5.0,-3.9,-3.9)$ for
$\kappa=1,~10^3$.

If field amplification is negligible in either shock, but  frozen-in magnetic
fields are radiatively dominant in the ejecta, there are three spectral
components
with fluences generally lower than in the shock turbulent growth case.
At cosmological distances, $E_o=10^{51},~D_{28}=1,\theta=10^{-1},~\eta=10^3,~
t_b=5\s$, we find $\log\Suxg\sim (-10.5,-10.5,-10.5),~(-7.6,-7.6,-6.2)$ for
$\kappa=1,~10^3$. In the galactic case, the radiative efficiency is much lower,
so
larger explosion energies are required, e.g.
$E_o=10^{43}\ergs,~\theta=10^{-1},~
\eta=10^2,~ t_b=5\s$, giving $\log\Suxg\sim (-7.8,-7.8,-7.8),~(-4.8,-4.8,-4.8)$
for $\kappa=1,~10^3$. The frozen-in cosmological (full line) and galactic (long
dashed
line) cases for $\ka=10^3$ are shown in Fig. 1.

Another possibility is that the fireball is so magnetically dominated that it
does not contain enough electrons to be accelerated by the reverse shock, or it
is matter dominated but the reverse shock is inefficient. Then, unless mixing
across the contact discontinuity occurs, the ejecta just act as a 'piston'. Two
spectral components can then arise if turbulent field growth and acceleration
occurs
in the blast wave. For the cosmological parameters above ($t_b=5\s$), the
fluences
are $\log\Suxg\sim (-7.3,-4.9,-4.7),~(-18.3,-15.8,-13.2)$ for $\kappa=
1,~10^3$,
while for galactic parameters ($E_o=10^{39},~\theta=10^{-1},\eta=3\times 10^1$)
one
expects $\log\Suxg\sim (-6.5,-6.5,-6.5),~(-6.4,-3.9,-3.9)$ for
$\kappa=1,~10^3$.
Two `piston model' spectra are shown in Fig. 1 for the cosmological case
($\eta=10^3,~
\ka=40,~ \xi=10^{-3},~t_b=5\s$, dot-dashed line) and for the galactic case
($E_o=10^{39}
\ergs,~\eta=10^2,~\ka=10^2,~\la=1,~t_b=0.2\s$, short dashed line).
\bsk
\ctl{\bf 5.~ Discussion}
\bsk
The detection of bursts (or even improved upper limits) in other wavebands
would aid greatly
in discriminating among various models. Moreover the present indications that
there
are two classes of classical gamma-ray bursts (Kouveliotou, \etal, 1993, Lamb,
Graziani and Smith, 1993) would be greatly strenghtened if these classses
turned out to
be distinctively different at lower frequencies. The results summarised here
relate
to models where the gamma-rays result from braking of relativistic ejecta by an
external medium. We originally applied these models to ``cosmological" bursts;
however,
a similar mechanism, with scaled-down parameters, may also be relevant to one
class
of ``galactic" bursts (Begelman, \Mesz and Rees, 1993).
The fluences  depend on $\eta,~n_o,~E_o,~D$ and $\theta$, and on the magnetic
field (though not in a simple manner), and the ratios of the fluences in the
three bands also depend in a complicated manner on the various parameters. The
numbers discusssed above, however, give an idea of some of the consequences of
various
physical assumptions. In most of the cases producing a detectable $\ga$-ray
emission,
a significant O/UV and X-ray emission is also predicted.

Matter dominated (loaded) models are radiatively efficient if strong fields
build
up behind both shocks, but lead to X-ray to $\ga$-ray ratios in excess of the
usually
invoked X-ray paucity requirement $L_x/L_\ga\siml 0.03$. This X-ray constraint
is better
satisfied by loaded models involving frozen-in magnetic fields in the ejecta,
which initially were close to equipartition, and where the reverse shock is
radiative (with or without a third component from blast wave IC). These models
can give $L_x/L_\ga \siml 0.03$ at cosmological distances (full lines, Fig. 1),
but not
at galactic disk distances, where all three fluences in the O/UV, X-ray and
$\ga$-ray bands
are comparable (long-dashed line, Fig. 1).

Magnetically dominated (unloaded) models, or loaded models with an inefficient
reverse
shock, can give ratios $L_x/L_\ga \siml 0.03$ in the ``piston" case, where only
the blast
wave radiates, but these models are less satisfactory in some other respects.
For
cosmological distances, $\kappa=4\times 10^1$ satisfies the X-ray paucity
criterion above
(dot-dashed line, Fig. 1), but the fluences are several orders of magnitude
lower than in
previous models. For galactic disk distances, $\kappa=10^3$ fits the X-ray
paucity spectrum
as written above (short dashed line, Fig. 1), but the duration
is $t_b\sim 0.2\s$, unless an extra free parameter is introduced for the
initial energy
deposition timescale.  On the other hand, if the blast wave does not radiate
(no
turbulent field growth) but local acceleration or mixing causes the reverse
shock to
be an efficient radiator, the cosmological case satisfies the X-ray paucity
constraint,
but the galactic case does not.

It may be possible to distinguish between the models above which satisfy the
X-ray paucity constraint, if positive measurements become available of the O/UV
fluences. Even if future observations brought about modifications of this
constraint
in some cases, predictions can be made about the expected ratio of fluence at
various
wavelengths.  The O/UV fluences in the (loaded or unloaded) frozen-in
cosmological
reverse shock case are predicted to be higher, comparable in energy to the
X-ray
fluence, or $\sim 1.5-2$ orders of magnitude below the $\ga$-ray fluence. On
the
other hand the unloaded magnetic piston or the matter-dominated piston models
predict an
O/UV fluence which is $\sim 1.5$ orders of magnitude below the X-ray fluence,
or $\sim
3.5-4.$ orders of magnitude below the $\ga$-ray fluence.
\bsk
{\it Acknowledgements:} We are grateful to NASA (NAGW-1522) and to the
Royal Society for support.
\bsk
\ctl{\bf References}
\ssk
\ref Band, D., \etal, 1993, Ap.J., 413, 281
\ref Begelman, M.C., \Mesz, P. and Rees, M.J., 1993, M.N.R.A.S., in press
\ref Blandford, R. and Eichler, D., 1987, Phys. Rep., 154, No.1, 1.
\ref Hartmann, D., \etal, 1993, to appear in {\it High Energy Astrophysics},
  J. Matthews, ed. (World Scientific).
\ref Hudec, R., \etal, 1992, in {\it Gamma-ray Bursts}, A.I.P. Conf.Proc. 265,
  W. Paciesas and Fishman, G., eds. (A.I.P., New York), p. 323
\ref Hoshino, M., Arons, J., Gallant, Y. and Langdon, B., 1992, Ap.J., 390, 454
\ref Hurley, K., 1989, in {\it 14th Texas Symp. Relat. Astrophysics, Ann. New
York Acad.
  Sci.}, 571, 442.
\ref Katz, J.I., 1993, Ap.J., in press
\ref Kouveliotou, C., \etal, 1993, Ap.J.(Letters), 413, L101
\ref Lamb, D.Q., Graziani, C. and Smith, I., 1993, Ap.J.(Lett.), 413, L11
\ref Laros, J.G., \etal, 1984, Ap.J., 286, 681
\ref \Mesz, P. and Rees, M.J., 1993, Ap.J., 405, 278
\ref \Mesz, P., Laguna, P. and Rees, M.J., 1993, Ap.J., in press
\ref Moskalenko, E.I., \etal, 1992, in {\it Gamma-ray Bursts}, C. Ho, R.
Epstein
  and E. Fenimore, eds. (Cambridge U.P.), p.127
\ref Murakami, T., \etal, 1992, in {\it Gamma-ray Bursts}, C. Ho, R. Epstein
  and E. Fenimore, eds. (Cambridge U.P.), p.239
\ref Narayan, R., \Pacz, B. and Piran, T., 1992, Ap.J.(Lett.), 395, L83
\ref \Pacz, B. and Rhoades, J., 1993, preprint
\ref Rees, M.J. and \Mesz, P., 1992, M.N.R.A.S., 258, 41P
\ref Ricker, G., \etal, 1992, in {\it Gamma-ray Bursts}, C. Ho, R. Epstein
  and E. Fenimore, eds. (Cambridge U.P.), p. 288
\ref Ryu, D. and Vishniac, E., 1991, Ap.J., 368, 411.
\ref Schaefer, B., \etal, 1992, Ap.J.(Lett.), 393, L51
\bsk
%
Fig. 1.-  Power per decade spectra for bursts with
$n_o=1\cmcui,~\theta=10^{-1}$.
1) Solid line: cosmological, frozen-in field, $E_o=10^{51}\ergs,~\eta=10^3,~
\kappa=10^3,~\xi=1,~t_b=5\s$. 2) Dot-dashes: cosmological, `piston',
$E_o=10^{51}\ergs,~\eta=10^3,~\kappa=40,~\xi=10^{-3},~t_b=5\s$. 3) Short
Dashes: galactic,
`piston', $E_o=10^{39}\ergs,~\eta=10^2,~\kappa=10^3,~\lambda=1,~t_b=0.2\s$.
4) Long Dashes: galactic, frozen-in field,
$E_o=10^{43}\ergs,~\eta=10^2,~\kappa=10^3,~\xi=1,~
t_b=5\s$. The first three satisfy the X-ray paucity constraint, but in 3) only
for
$t_b\siml 1\s$, while 4) overproduces X-rays.
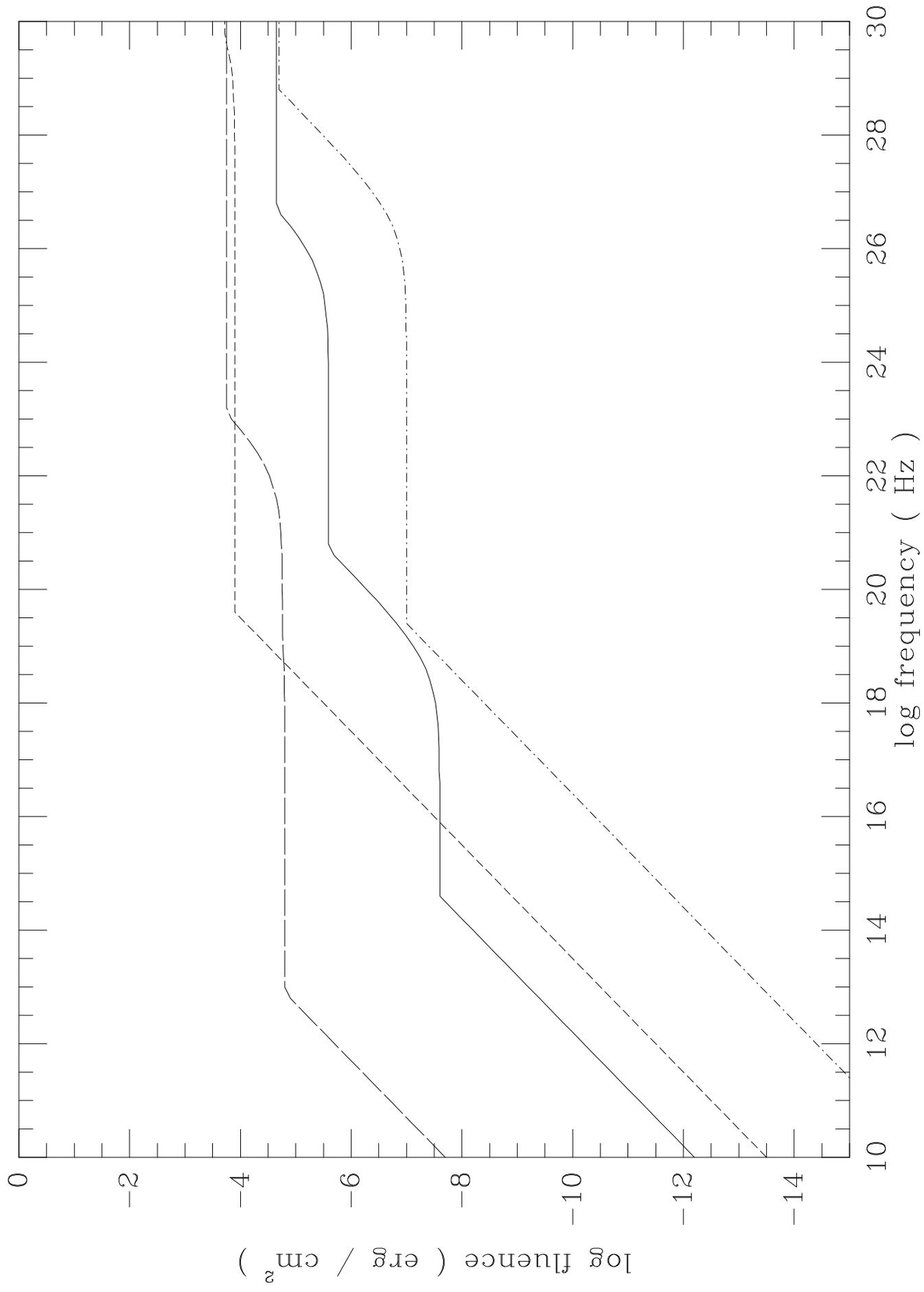
\end